\documentclass{ws-procs975x65}            
\begin{document}
\title{New constraints on quantum foam models from X-ray and gamma-ray observations of distant quasars}

\author{Eric S. Perlman}


\address{Dept. of Physics \& Space Sciences, Florida Institute of Technology,  Melbourne, FL  32901, USA\\ eperlman@fit.edu}

\author{Saul A. Rappaport}

\address{Dept. of Physics and Kavli Institute for Astrophysics \& Space Research\\ Massachusetts Institute of Technology, Cambridge, MA 02139, USA\\ sar@mit.edu}

\author{Y. Jack Ng, Wayne A. Christiansen}

\address{Dept. of Physics and Astronomy, University of North Carolina, Chapel Hill, NC  27599, USA \\  yjng@physics.unc.edu, wayne@physics.unc.edu }

\author{John DeVore}

\address{Visidyne, Inc.; jgdevore@cox.net}

\author{David Pooley}

\address{Dept. of Physics \& Astronomy, Trinity University, San Antonio, TX  78212, USA \\ dpooley@trinity.edu}

\begin{abstract}
Astronomical observations of distant quasars may be important to test models for quantum gravity, which posit Planck-scale
spatial uncertainties ('spacetime foam') that would produce phase fluctuations in the wavefront of radiation emitted by a source, which may accumulate over large path lengths. 
We show explicitly how wavefront distortions cause the image intensity to decay to the point where distant objects become undetectable if the accumulated path-length fluctuations become comparable to the wavelength of the radiation. We also reassess previous efforts in 
this area.
We use X-ray and gamma-ray observations to rule out several models of spacetime foam, including the interesting random-walk and
holographic models. 
\end{abstract}

\bodymatter	

\section{Introduction}
\label{sec:Intro}

Even at the minute scales of distance and duration examined with
increasingly discriminating instruments, spacetime still appears
to be smooth and structureless. However, a variety of models of 
quantum gravity posit that spacetime is, on Planck scales, subject to
quantum fluctuations. As such, the effect of quantum gravity 
on light propagation (if detected) can possibly reveal a coupling to vacuum 
states postulated by Inflation and String Theories.  
In particular, models 
\cite{ng03b}
 consistent with the ``Holographic Principle" 
 \cite{tho93, sus95,aha00}
predict that space-time foam 
may be detectable via intensity-degraded or blurred images of distant objects.  
While these models are not a direct test of the Holographic Principle itself, the 
success or failure of such models may provide important clues  to connect black 
hole physics with quantum gravity and information theory. 
\cite{haw75}

The fundamental idea is that,  if probed at a small enough scale, spacetime 
will appear complicated -- something akin in complexity to a turbulent
froth that Wheeler (1963)\cite{whe63} has 
dubbed ``quantum foam,'' also known as 
``spacetime foam.''  In models of quantum gravity, the ÒfoaminessÓ of spacetime 
is a consequence of the Energy Uncertainty Principle connecting the Planck mass 
and Planck time.  Thus, the 
detection of spacetime foam is important for constraining models of 
quantum gravity.   If a foamy structure is found, it would require
that space-time itself has a probabilistic, rather than deterministic nature.  As a 
result, the phases of photons emitted by a distant source would acquire a random 
component which increases with distance.  

A number of prior studies have explored the possible image degradation of distant 
astronomical objects due to the effects of spacetime foam. 
\cite{lie03,ng03b,rag03,chr06,CNFP, PNFC}
In particular, most of these focus on possible
image blurring of distant astronomical objects. We demonstrate that this previous approach 
was incomplete, and take a different approach, examining the possibility that spacetime foam 
might actually prevent the appearance of images altogether at sufficiently short wavelengths. 
Short-wavelength observations are particularly useful in constraining quantum 
gravity models since, in most models of quantum gravity, the path-length fluctuations 
and the corresponding phase fluctuations imparted to the wavefront of the radiation 
emitted by a distant source are given \cite{CNFP} by:
\begin{equation}
\label{eqn:phase}
\delta \phi \simeq 2 \pi \ell^{1-\alpha} \ell_P^{\alpha} / \lambda
\end{equation} 
where $\lambda$ is the 
wavelength one is observing, the parameter $\alpha \lesssim 1$ specifies different 
space-time foam models, and $\ell$ is the line-of-sight co-moving distance to the source.  

\section{Effects of Spacetime Foam on Astronomical Images}
\label{sec:phase_effects}

As discussed in the 
Introduction, there are good reasons to believe that spacetime foam would produce
small phase shifts in the wavefronts of light arriving at telescopes.
Eqn.~(\ref{eqn:phase}) points out that the path-length fluctuations  envisioned by models
of quantum gravity distort the wavefront emitted by cosmologically distant sources, 
imparting phase fluctuations.  The individual fluctuations are infinitesimally small, but 
depending on the model for quantum gravity being discussed, they may accumulate over 
long path lengths, perhaps to a point where their effects can be detected.  This is the
essence of our work.  To help understand what images of distant, unresolved sources
(e.g., quasars or gamma-ray bursts) might look like after propagating to Earth through a
space-time foam-induced 'phase screen',  we consider the Fourier
transform of $e^{i \Delta \phi(x,y)}$ over the coordinates $\{x,y\}$ of the entrance aperture.
This approach is particularly
helpful because the image formed will be simply the absolute square of this function.  
Moreover,  if the phase fluctuations in $\Delta \phi$ are assumed to be analytic, part of 
the problem can also be done analytically \cite{P15}, 
making simulations particularly easy to carry out.  

In particular, we find that as long as $\delta \phi_{\rm rms} \lesssim 0.6$ radians
(or $\delta \ell_{\rm rms} \lesssim 0.1 \lambda$) the Strehl ratio, which measures 
the ratio of the point spread function ('PSF') compared to an ideal PSF for the 
same optics, is to a good approximation, 
\begin{equation}
S \simeq e^{-\delta \phi_{\rm rms}^2} ~~.
\label{eqn:Strehl}
\end{equation}
Furthermore, if these phase shifts are distributed randomly over the aperture (unlike 
the case of phase shifts associated with well-known aberrations, such as coma,
astigmatism, etc.) then the {\em shape} of the PSF, after the inclusion of the phase shifts
due to the spacetime foam is basically unchanged, except for a progressive decrease 
in $S$ with increasing $\delta \phi_{\rm rms}$.  

\begin{figure}
\centering
%
%
%
\includegraphics[width=1.00\columnwidth]{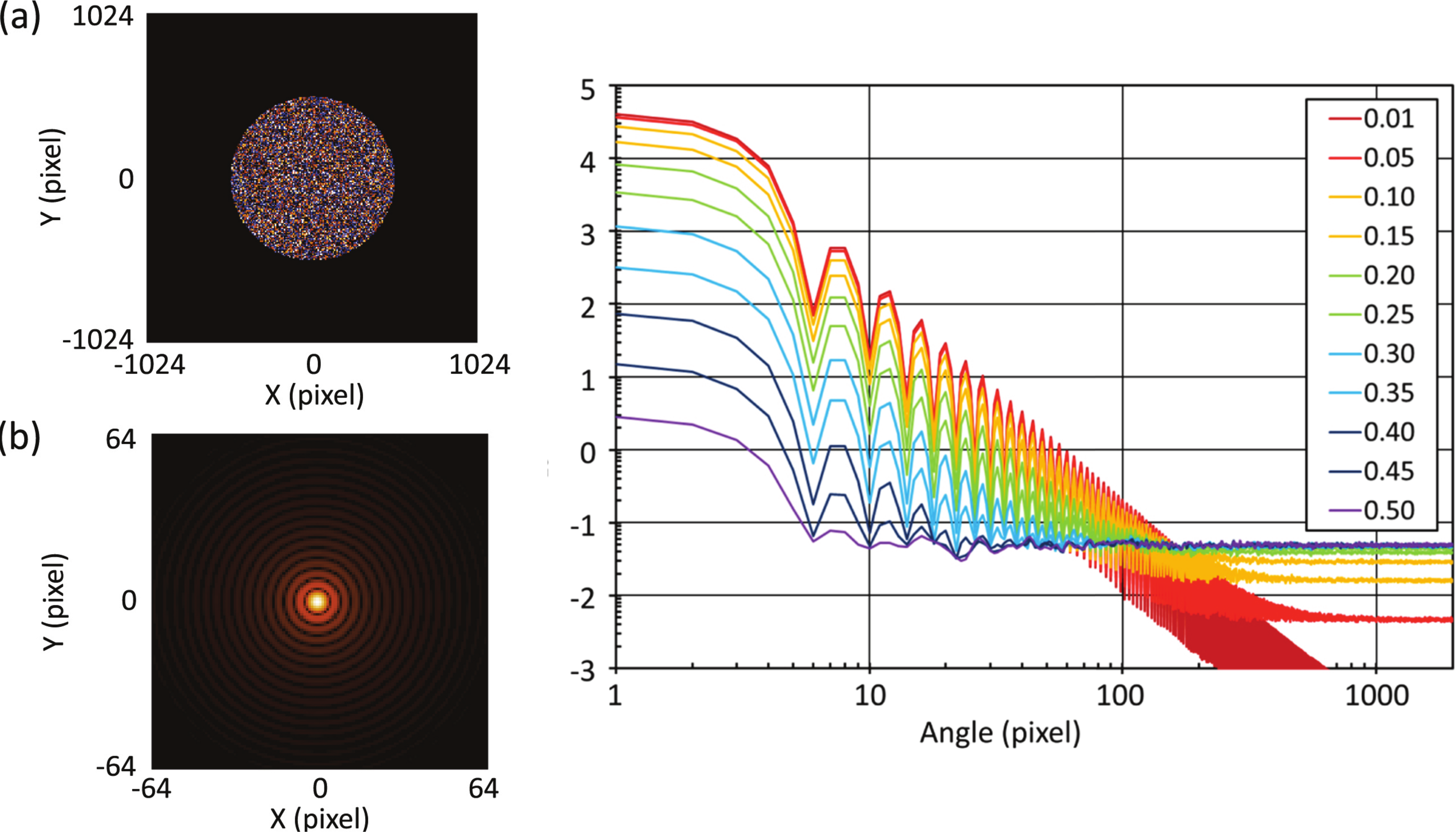}
\caption{At left, an illustrative example of our numerical simulations of a point spread function 
that has been affected by a Gaussian random field of phase shifts over the aperture.  The top left panel shows the circular aperture with Gaussian phase shifts of rms amplitude $ 0.03 \lambda$.  The bottom leftt panel  shows the inner $128 \times 128$ pixels of the absolute square 
of the Fourier transform of the aperture function, using a 1/4-power law color palette.  The right-hand panel shows a sequence
of radial profiles of the numerically computed PSFs for 
rms phase shifts ranging from 0.01 $\lambda$ to 0.5 $\lambda$, as indicated by the color coding.   Note how the {\em shape} of the PSF for small angles is nearly unchanged
until it plateaus into the background.} 
\label{fig:PSF2}
\end{figure}

We carried out numerical simulations utilizing various random fields $\Delta \phi(x,y)$,
including Gaussian, linear, and exponential.  We considered a large range of rms values and 
different correlation lengths within the aperture.  
Fig.~\ref{fig:PSF2} illustrates these simulations, and shows a sequence of the simulated 
%
PSFs, in the form of radial profiles,
for a range of increasing amplitudes of random phase fluctuations.  As can
be seen, there are three major effects: (i) the peak of the 
PSF is decreased; (ii) beyond a certain radial distance, the PSF reaches a noise 
plateau that can be interpreted as an indication of the partial de-correlation of the wave 
caused by increasing phase fluctuations; 
and (iii) in between, the shape (including the slope, intensity ratios of Airy rings, etc.) 
of the PSF is {\it unchanged} by the increasing phase fluctuations. The self-similar invariance of 
the PSF shape (aside from the appearance of the noise plateau) contradicts the expectation 
from previous work 
\cite{lie03,ng03a, ng03b,rag03,chr06,CNFP, PNFC,ste07,ste15,tam}
that phase 
fluctuations could broaden images of a distant quasars. 
In contrast, we now find that while the images are essentially unaffected, for 
sufficiently large amplitude phase fluctuations (e.g., $\delta \ell/\lambda \gtrsim \pi$) the entire 
central peak disappears and the image is undetectable.

As Figure \ref{fig:PSF2}  shows, the overall PSF shape for a cosmologically distant source, and
the slope of its decline, will be nearly unchanged until the phase differences imposed by spacetime 
foam approach $\sim$$\pi$ radians, at which point the profile just merges into the background 
noise floor.   At this point, the quasar intensity would basically be 
degraded to the point where it would no longer be detected.  This forces us to re-conceptualize 
how one might constrain models for quantum gravity, and particularly the $\alpha$-models.  By 
inverting our analytical model of the PSF, we find that 
\begin{equation}
\label{eqn:constraint}
\alpha > \frac{\ln(\pi \ell/\lambda)}{\ln(\ell/\ell_P)},~~
\end{equation} 
{\noindent where we have required a phase dispersion $\delta \phi_{rms} = 2$ radians, 
corresponding to the location where the Strehl ratio falls to $\sim$2\% of its full value. 
We show in Fig.~\ref{fig:alpha} a plot of the limit that can be set on the parameter $\alpha$ 
as a function of measurement wavelength, for four different values of comoving distance.  
The result is an essentially universal constraint that can be
set simply by the detection of distant quasars as a function of the observing wavelength.
This more rigorous understanding has significant effects on the constraints one can
set on $\alpha$ using observations in any given waveband.  While it loosens the constraints set by 
optical observations to  $\alpha> 0.53$, contrary to 
previous works (including our own), i.e., ruling out the random walk 
model, but not coming close to the parameter space required for the holographic model.  
Another way to think of this constraint\cite{P15} is that for any given wavelength, $\alpha$-models 
predict that there is a maximum distance, beyond which it would be impossible to detect a 
source.

The simulations we have done have profound implications for constraining the 
spacetime foam parameter $\alpha$.   Equation (1)
shows that for a given source distance, $\ell$, the rms phase shifts over the wavefront
are proportional to $\lambda^{-1}$.  This opens up the possibility of using X-ray and 
gamma-ray observations to set the tightest constraints yet.  The constraints produced in a 
given band are symbolized in Fig. \ref{fig:alpha} by vertical lines 
that denote optical (5000 \AA~ wavelength or 2.48 eV photon energy), X-ray (5 keV), GeV and TeV 
photons.  These represent the energies where observations of distant quasars by the  {\it Chandra} X-ray Observatory, 
{\it Fermi} Gamma-ray Space Telescope, and the VERITAS telescope array, 
\cite{P15} all of which show well-resolved images, may be used to constrain $\alpha$.
The constraints thus produced (Fig. \ref{fig:alpha}) are lower limits to $\alpha$ produced by the 
mere observation
of an image (whether diffraction limited or not!) of a cosmologically distant quasar.


\begin{figure}
\centering
{\includegraphics[width=0.9\columnwidth]{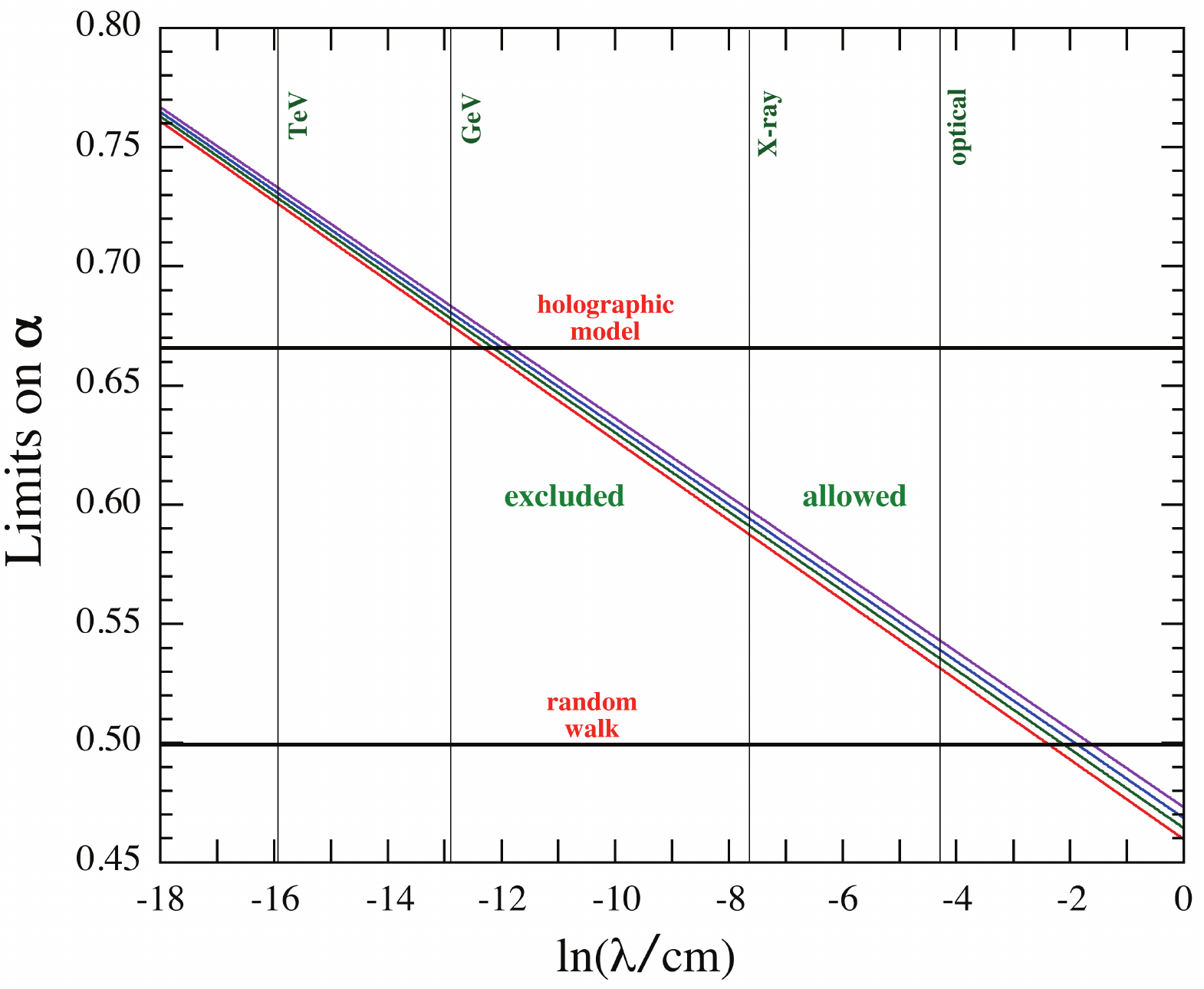}}
\caption{Constraints on the \ parameter $\alpha$, for four different comoving distances
to the object, respectively 300 Mpc ($z\approx 0.07$; red curve), 1 Gpc ($z\approx0.25$; green), 
3 Gpc ($z\approx 1$; blue) and 10 Gpc ($z\approx 12$; purple).  The two horizontal  refer 
to the holographic and random-walk models, respectively, as labeled.  The vertical dashed lines 
represent the optical (5000 \AA), X-ray (5 keV), GeV and TeV wavebands.  As astronomical images
betray no evidence of cosmic phase fluctuations that might be due to spacetime foam, the region of
parameter space excluded by observations in each band lies below the curves. 
For any given wavelength, $\lambda$, images will not propagate for values of $\alpha$ below the 
various lines corresponding to different comoving distances. } 
\bigskip
\label{fig:alpha}
\end{figure}
%
%

\section{Summary and Conclusions}
\label{sec:summary}

According to the simulations discussed here, it would seem that the very {\em existence} of distant
astronomical images can be used to put significant constraints on models of spacetime foam.
Perhaps the strongest constraints of all now come from the detection of large numbers 
of cosmologically distant sources -- mostly blazars -- in the $\gamma$-rays.  
These detections limit $\alpha$ to values higher than 0.67 and 0.72, at GeV and TeV energies, 
respectively.  
This strongly disfavors, if not completely rules out, the holographic model.

There are, however, a number of caveats to our idea for constraining $\alpha$-models of spacetime foam.
In particular, 
as pointed out by Stefano Liberati at this meeting, GeV and even TeV gamma-rays have 
wavelengths that are vastly larger than the Planck scale.  It is possible that photons propagate 
on an averaged space-time, so that their geodetics would be smooth, not noticeably affected by 
spacetime foam effects.  
Another possibility along these same lines 
\cite{coumbe}
is that time dilates as a 
function of distance scale, so that independent of the value of $\alpha$, geodetics would be 
unaffected by space-time foam effects and no phase dispersion would be expected, regardless of
the wavelength of the photon.  The latter proposal carries with it also an interesting prediction that 
the speed of light is energy independent.
At this time it is difficult for us to see how to simulate such an effect.  

In this work we have considered the instantaneous fluctuations in the distance between the location of the emission and a given point on the telescope aperture. Perhaps one should average over both the huge number of Planck timescales during the time it takes light to propagate through the telescope system, and over the equally large number of Planck squares across the detector aperture. It is then possible that the fluctuations we have been calculating vanish, but at the moment we have no formalism for carrying out such averages.

Finally, we should recall that the 
spacetime foam model parametrized by $\alpha = 2/3$, as formulated, 
\cite{ng94,ng95}
is called the `holographic model' only because it is consistent 
\cite{ng03b}
with the holographic
principle; the demise of the model may {\it not} necessarily imply the demise
of the principle since it is conceivable that the correct spacetime 
foam model associated with the holographic principle can take on a 
different and more subtle form than that which can be
given by $\delta \ell \approx \ell^{1/3} \ell_P^{2/3}$.  
It is important to be clear: what we are ruling out (subject to the
caveats mentioned above) are the models with
$\alpha < 0.72$ for the spacetime foam models that can be categorized
according to $\delta \ell \approx \ell^{1 - \alpha} \ell_P^{\alpha}$. 


\end{document}